\documentclass{PoS}

\title{Testing a generalized cooling procedure in the complex Langevin simulation of chiral Random Matrix Theory}

\usepackage{amssymb,amsmath}
\usepackage{subfigure}
\graphicspath{{./Fig/}}

\ShortTitle{Testing a generalized cooling procedure 
in the complex Langevin simulation...}

\author{\speaker{Keitaro Nagata}%
         \thanks{KEK-CP-327,KEK-TH-1874}\\
KEK Theory Center,
High Energy Accelerator Research Organization, Tsukuba 305-0801, Japan\\
        E-mail: \email{knagata@post.kek.jp}}

\author{Jun Nishimura\\
KEK Theory Center,
High Energy Accelerator Research Organization, Tsukuba 305-0801, Japan\\
Department of Particle and Nuclear Physics, 
School of High Energy Accelerator Science, 
Graduate University for Advanced Studies (SOKENDAI), 
Tsukuba 305-0801, Japan \\
        E-mail: \email{jnishi@post.kek.jp}}

\author{Shinji Shimasaki\\
KEK Theory Center,
High Energy Accelerator Research Organization, Tsukuba 305-0801, Japan\\
Research and Education Center for Natural Sciences, Keio University,
Hiyoshi 4-1-1, Yokohama, Kanagawa 223-8521, Japan \\
        E-mail: \email{simasaki@post.kek.jp}}

\usepackage{graphicx}
\usepackage{color}

\newcommand{\SLNC}[1]{{\rm SL}(#1, \mathbb{C})}
\newcommand{\GLNC}[1]{{\rm GL}(#1, \mathbb{C})}
\newcommand{\SUN}[1]{{\rm SU}(N)}
\newcommand{\UN}[1]{{\rm U}(N)}

\newcommand {\ee}{\mbox{e}}

\abstract{
The complex Langevin method has been attracting much attention 
as a solution to the sign problem since the method was shown
to work in finite density QCD in the deconfined phase
by using the so-called gauge cooling procedure. 
Whether it works also in the confined phase with light quarks is
still an open question, though. In order to shed light on this question,
we apply the method to the chiral Random Matrix Theory,
which describes the epsilon regime of finite density QCD.
Earlier works reported that a naive implementation of the method
fails to reproduce the known exact results and that the problem 
can be solved by choosing a suitable coordinate.
In this work we stick to the naive implementation, and show that a generalized 
gauge cooling procedure can be used to avoid the problem.}

\FullConference{The 33rd International Symposium on Lattice Field Theory\\
                 14 -18 July  2015\\
                 Kobe International Conference Center, Kobe, Japan}

\begin{document}

\section{Introduction}

The complex Langevin method (CLM) is 
a promising candidate of solutions to the sign problem,
which occurs in studying various interesting systems 
such as QCD at finite density
by Monte Carlo methods.
It may be viewed as an extension of 
the Langevin method or the stochastic quantization for real action systems, 
which updates dynamical variables using the Langevin equation \cite{Damgaard:1987rr}. 
Since the Langevin method does not rely on the probability interpretation
unlike Monte Carlo methods, 
it may be generalized to the case with
a complex action \cite{Parisi:1984cs,Klauder:1983sp}.

When one solves the Langevin equation in the case of complex action, 
the complex drift term inevitably makes
the dynamical variables of the system evolve into complex values
even if they are originally real.
For the method to work nevertheless, the action and the observables have to be 
extended to holomorphic functions of 
the complexified dynamical variables.
In addition to this, it was found that
the probability distribution of the complexified dynamical variables
should have a fast fall-off
in the imaginary directions \cite{Aarts:2009uq,Aarts:2011ax}. 

In the application of CLM to QCD 
at finite density,
the drift term makes the SU(3) link variables evolve into $\SLNC{3}$ matrices. 
It can then happen that
the link variables make long excursions in the non-compact directions 
with large probability, which causes convergence to a wrong result. 
In order to avoid this problem,
the so-called gauge cooling was proposed \cite{Seiler:2012wz}.
This amounts to performing a complexified $\SLNC{3}$ gauge transformation 
after each Langevin step in such a way that one can reduce
the unitarity norm, which measures the deviation of the link variables 
from unitary matrices. 
Using this technique, the CLM was shown to work for QCD 
in the heavy dense limit \cite{Seiler:2012wz}
or at sufficiently high temperature \cite{Sexty:2013ica,Fodor:2015doa}.
Recently an explicit justification of the CLM 
including the gauge cooling procedure
was given by the present authors \cite{Nagata:2015uga}.
Since the advent of this new technique, 
the CLM has been attracting much attention; 
see e.g., 
\cite{Tsutsui:2015tua,%
Bloch:2015coa,Sinclair:2015kva,Aarts:2015hnb,Aarts:2015yuz,Hayata:2015lzj}.

In fact, when one tries to apply 
the CLM to finite density QCD at low temperature and
with light quarks, another problem is anticipated 
to occur \cite{Mollgaard:2013qra,Greensite:2014cxa}.
The nonzero quark chemical potential breaks 
the anti-Hermiticity of the Dirac operator $D$, 
and causes the broadening of its eigenvalue distribution.
For quark chemical potential larger than a certain value,
the eigenvalue distribution of $(D+m)$ with $m$ being the quark mass
becomes nonzero at the origin,
and the diverging drift term in the Langevin equation 
causes convergence to a wrong result \cite{Nishimura:2015pba}. 
The gauge cooling with the unitary norm does not seem to cure this problem
because the eigenvalue distribution is broadened
at nonzero quark chemical potential even if the link variables are unitary. 

The purpose of this work is to develop a method 
which solves this fermionic singular-drift problem. 
For that, we extend the idea of gauge cooling by 
introducing new types of norms
written in terms of the Dirac operator,
which are sensitive to the problematic features of the eigenvalue distribution.
We test the method in the chiral Random Matrix Theory (chRMT), 
for which naive implementation of the Langevin simulation fails 
in the light quark regime \cite{Mollgaard:2013qra}.
While the chRMT does not have a gauge symmetry,
it has U($N$) global symmetries, which can be used
to cure the problem \cite{Nagata:2015uga}.
In ref.~\cite{Mollgaard:2014mga} 
the same problem was solved by
using the polar coordinate, instead of Cartesian,
for the dynamical variables.
In this work, however, we stick to the Cartesian coordinate and
test the proposed method,
which can be applied to QCD and other cases in a straightforward manner.
We will show that reducing the new types of norm by applying
the cooling procedure after each Langevin step can indeed 
make the Langevin simulation converge to the correct results.

\section{Chiral Random Matrix Theory}
We consider a chRMT
for $N_{\rm f}$ quarks with degenerate mass $m$ and the chemical potential $\mu$, 
whose partition function is given by 
\begin{align}
Z &= \int d\Phi_1 d\Phi_2
\, [\det (D+m)]^{N_{\rm f}} \, \ee^{-S_{\rm b}} \ .
\label{crmt}
\end{align}
Here the dynamical variables are $N\times N$ general 
complex matrices\footnote{The topological index $\nu$ can be introduced 
in our analysis 
by making the matrices $\Phi_k$ and $\Psi_k$
rectangular as $N\times (N+\nu)$ and  $(N+\nu)\times N$,
respectively. The complexified symmetry (\ref{Eq:2015Apr14eq1}) with
$g \in \GLNC{N}$ and $h \in \GLNC{N+\nu}$ can be used for gauge cooling
in that case.}
$\Phi_k$ ($k=1,2$),
and the bosonic action $S_{\rm b}$ is given by 
\begin{align}
S_{\rm b} &=2N \, \sum_{k=1}^2 {\rm Tr} \, (\Psi_k \Phi_k) \ ,
\label{Eq:2015Apr09eq2}
\end{align}
where we have defined $\Psi_k$ ($k=1,2$) by
\begin{align}
\Psi_k &= (\Phi_k) ^\dag
\label{def of Phipm}
\end{align}
for later convenience.
The Dirac operator $D$ is defined by 
\begin{align}
D& =\left( \begin{matrix}
0 & X \\
Y & 0
\end{matrix} \right) \ ,  \\
X &= \ee^{\mu} \Phi_1 + \ee^{-\mu} \Phi_2 
\ , \\
Y &= -\ee^{-\mu} \Psi_1 - \ee^\mu \Psi_2 \ .
\end{align}
At $\mu=0$, the Dirac operator is anti-Hermitian $D^\dagger = - D$, 
from which it follows that
its eigenvalues are purely imaginary. 
At $\mu\neq 0$, this is not the case any more, 
and the eigenvalues become general complex numbers. 
%
As interesting observables, we consider the chiral condensate $\Sigma$ 
and the baryon number density $n_{\rm B}$ defined, respectively, by
\begin{align}
\Sigma = \frac{1}{N} \frac{\partial}{\partial m} \log Z \ ,
\quad \quad
%
n_{\rm B} = \frac{1}{N} \frac{\partial}{\partial \mu} \log Z \ .
\label{observables}
\end{align}
%
Note that the theory \eqref{crmt} as well as the observables
(\ref{observables})
is invariant under the transformation 
\begin{align}
\Phi_k \to \Phi_k ' = g \, \Phi_k \, h^{-1} \ ,
\quad\quad
\Psi_k \to \Psi_k ' = h \, \Psi_k \, g^{-1} \ ,
\label{Eq:2015Apr14eq1}
\end{align}
where $g, h \in \UN{N}$. 
%

\section{New gauge cooling for solving the fermionic singular-drift problem}
\label{sec:cooling}

In the CLM, real variables have to be extended to complex variables
in a holomorphic manner.
This amounts
to removing the constraint (\ref{def of Phipm})
and treating $\Phi_k$ and $\Psi_k$ ($k=1,2$)
as independent variables.
Then the symmetry is naturally complexified, and the invariance
under (\ref{Eq:2015Apr14eq1}) holds for $g,h \in \GLNC{N}$.
The ``gauge cooling'' can be introduced as a procedure of
making a complexified symmetry transformation 
after each Langevin step so that the norm we define below is reduced.

Originally the gauge cooling was proposed
to avoid the problem due to the long excursion  
in the imaginary directions \cite{Aarts:2009uq,Aarts:2011ax},
and for that purpose, the unitarity norm was
introduced in QCD \cite{Seiler:2012wz}.
Analogously, in the chRMT we consider the norm
\begin{align}
\mathcal N_{\rm h} = \frac{1}{N} \sum_{k=1}^2 
{\rm Tr} \, 
\left[ \Big\{ \Psi_k - (\Phi_k)^\dagger \Big\} ^\dagger 
\Big\{ \Psi_k - (\Phi_k)^\dagger \Big\} \right] \ .
\label{hermiticity-norm}
\end{align}

In ref.~\cite{Mollgaard:2013qra},
convergence to a wrong result was observed
at small quark mass.
It is understood that 
this problem is caused by the diverging drift term
due to the small eigenvalues of $(D+m)$ \cite{Nishimura:2015pba}.
According to this understanding, we should be able
to cure the problem by using the gauge cooling
in such a way that small eigenvalues of $(D+m)$ are suppressed.
For that purpose, we introduce new types of norms. 
The first one is given by 
\begin{align}
\mathcal N_1 = \frac{1}{N} \, {\rm Tr} \, 
\Big[ (X+Y^\dagger)^\dagger (X+Y^\dagger) \Big]
\ ,
\end{align} 
which measures the violation of the anti-Hermiticity of $D$.
Reducing this norm makes the eigenvalue distribution of $D$ 
narrower in the real direction, and thus 
the appearance of small eigenvalues of $(D+m)$ can be suppressed.
The second one is defined by 
\begin{align}
\mathcal N_{2} = \sum_{a=1}^{n_{\rm ev}} \ee^{-\xi \alpha_a } \ , 
\label{normtype2}
\end{align}
where $\xi$ is a real parameter
and $\alpha_a$ are the real positive eigenvalues of $M^\dagger M$ with $M=D+m$.
The sum in (\ref{normtype2}) is taken over the $n_{\rm ev}$ 
smallest eigenvalues of $M^\dagger M$. 
Since $\alpha_a \ge \delta$ implies $|\lambda_a|^2 \ge \delta$,
where $\lambda_a$ are the eigenvalues of $M$,
reducing the norm $\mathcal N_2$
suppresses the appearance of small $\lambda_a$.

In order to avoid the long-excursion problem and 
the fermionic singular-drift problem
at the same time,
we consider a linear combination of the norms $\mathcal N_{\rm h}$
and $\mathcal N_{i}$ given by
\begin{align}
\mathcal N_i (r) = r \mathcal N_{\rm h} + (1-r) \mathcal N_i \ , 
\quad \quad 0 \le r \le 1 \ , 
\label{N-linear-combi}
\end{align}
where 
$i=1,2$ and $r$ is a tunable parameter.
Note that the eigenvalue distribution of $M$ is actually
invariant under the $\GLNC{N}$ transformation, and yet
it is affected by the gauge-cooling procedure,
which 
is possible because the noise term respects only $\UN{N}$.


\section{Results}
\label{sec:result}

We have studied the chRMT (\ref{crmt}) for
$N=30$,  $\sqrt{N} \mu = 2$ and $N_{\rm f}=2$ by
solving the discretized complex Langevin equation.
We have made $100000$ steps of time-evolution
with the step size $\Delta t = 5\times 10^{-5}$.
This step size was small enough to avoid the runaway problem. 
The initial 20000 steps were discarded for thermalization,
and the observables were measured every 200 steps.

The parameter $r$ in (\ref{N-linear-combi})
is chosen to be $r=0$ for $\mathcal N_1$
and $r=0.01$ for $\mathcal N_2$.
These small values of $r$ turned out to be good
enough to control (\ref{hermiticity-norm})
since the long excursion in the imaginary direction 
is suppressed by the bosonic action (\ref{Eq:2015Apr09eq2})
in the present case. 
The situation might be different in QCD. 
The parameters
used in (\ref{normtype2}) for the $\mathcal N_2$ norm
are chosen as $\xi=300$ and $n_{\rm ev}=2$. 

The gauge cooling procedure is performed similarly 
to the original proposal \cite{Seiler:2012wz}.
We calculate the gradient of the norm
with respect to the complexified transformation,
and choose the magnitude of the transformation parameter
in such a way that the norm is maximally reduced.
This procedure is repeated 10 times after each Langevin step.

\begin{figure}[htbp] 
\begin{center}
\includegraphics[width=6cm]{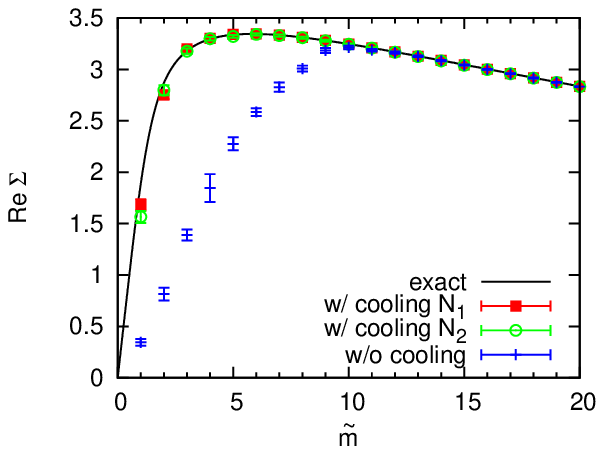}
\includegraphics[width=6cm]{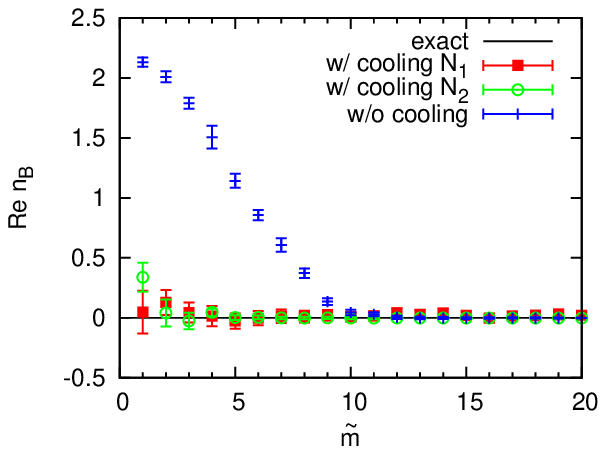}
\end{center}
\caption{The chiral condensate (Left) and the baryon number density
(Right) obtained by the CLM with or without gauge cooling
are plotted against $\tilde{m}=m N$.
The solid lines represent the exact results.}
\label{Fig:2015Jul29Fig1}
\end{figure}

In Fig.~\ref{Fig:2015Jul29Fig1} we show our results for
the chiral condensate and the baryon number density 
obtained by the CLM with or without gauge cooling.
The CLM without gauge cooling fails for $m N \lesssim 10$
as was reported in ref.~\cite{Mollgaard:2013qra},
whereas the CLM with gauge cooling using new types of norm successfully 
reproduces the exact results even in the  light quark regime. 

\begin{figure}[htbp] 
\begin{center}
\subfigure[without gauge cooling]
{\includegraphics[width=6cm]{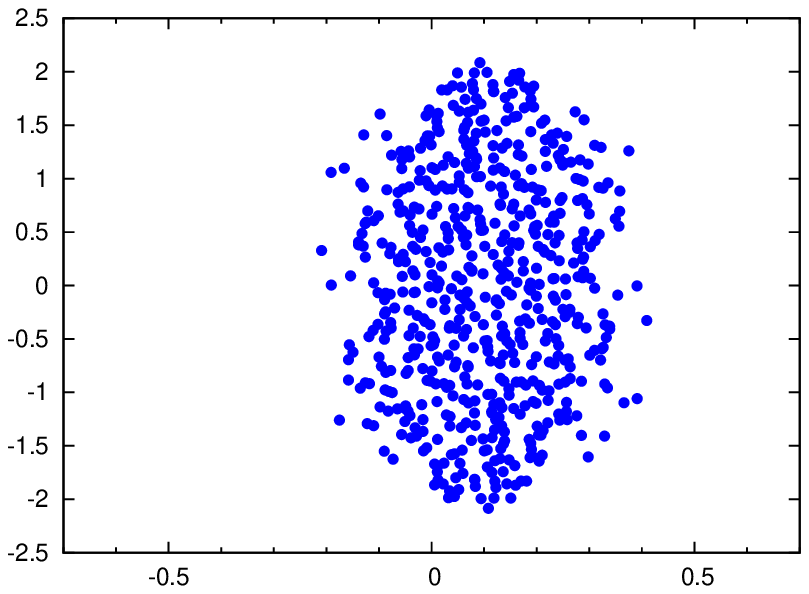}}
\hspace{0cm}
\subfigure[with gauge cooling using the $\mathcal N_1$ norm]
{\includegraphics[width=6cm]{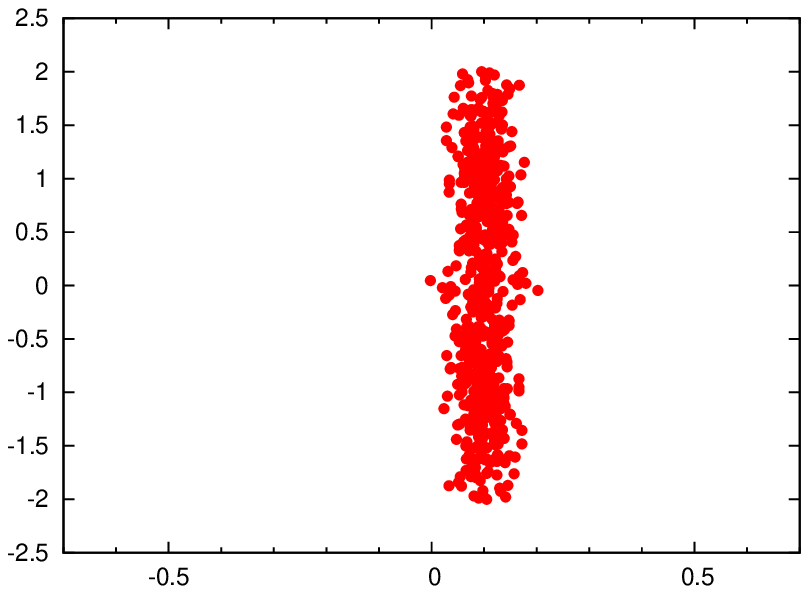}} \\
\subfigure[with gauge cooling using the $\mathcal N_2$ norm]
{\includegraphics[width=6cm]{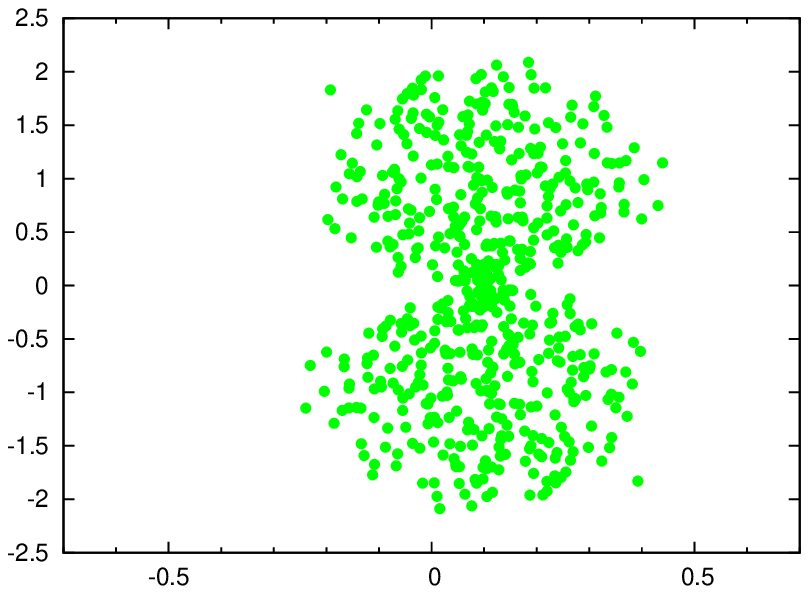}}
\hspace{0cm}
\caption{The eigenvalue distribution of $D+m$ obtained by the CLM in each case.
The data points represent the eigenvalues 
obtained from the last 10 configurations with intervals of 200 steps.
}
\end{center}
\label{Fig:eigenvalue distribution}
\end{figure}
In Fig.~2 we show the eigenvalue distribution of $(D+m)$ obtained with
$\tilde{m}= m N = 3$, for which the CLM works only 
with the new gauge cooling.
Let us recall that the eigenvalues 
close to zero
make the drift term large in the Langevin equation,
which invalidates the argument for justifying the CLM \cite{Nishimura:2015pba}.
Without gauge cooling, we do find that the origin
is covered with the cloud of eigenvalues completely,
whereas with gauge cooling,
that is somehow avoided.
It is interesting that
the problem is avoided in different ways for the two types of norm.
The gauge cooling with ${\cal N}_1$ makes the eigenvalue distribution
narrower in the real direction, 
whereas the gauge cooling with ${\cal N}_2$ repels the eigenvalues from
the vicinity of the origin. 

Note that the eigenvalue distribution of $(D+m)$ is \emph{not} 
a holomorphic quantity,
and hence the results shown in Fig.~2 do not represent
the eigenvalue distribution of $(D+m)$ that can be defined in the chRMT,
which is actually complex. (See, for instance, ref.~\cite{Splittorff:2014zca}.)
Therefore, it is possible that the eigenvalue distribution obtained by the CLM
depends on the norm chosen for the gauge cooling even if the method is working.
On the other hand, the chiral condensate
can be expressed in terms of the eigenvalue distribution obtained 
by the CLM \cite{Splittorff:2014zca},
and yet it is a holomorphic quantity.
Such a quantity should be universal and should not depend
on the norm chosen for the gauge cooling as long as the method is working.
Details will be reported in 
our forthcoming paper \cite{Nagata:2015future}.
\section{Summary}

In this work we have proposed new types of norm 
written in terms of the Dirac operator
that can be used in gauge cooling 
to overcome the fermionic singular-drift problem in the CLM. 
We have demonstrated that the gauge cooling with the proposed norms 
makes the method work
even in the small quark mass region.
While the critical quark mass, below which we cannot study by the CLM,
still exists and increases with the chemical potential \cite{Nagata:2015future},
the range of applicability of the CLM is greatly enlarged by the new gauge cooling.
The proposed method
can be 
applied to finite density QCD in a straightforward manner.
It is interesting to explore the parameter region accessible by the method.

\acknowledgments
K.~N.\ was supported by JSPS Grants-in-Aid for Scientific Research (Kakenhi) Grants 
No.\ 00586901, MEXT SPIRE and JICFuS. 
The work of J.~N.\ was supported in part by 
Grant-in-Aid for Scientific Research (No.\ 23244057) 
from Japan Society for the Promotion of Science.


\begin{thebibliography}{10}

\bibitem{Damgaard:1987rr}
P.~H. Damgaard and H.~Huffel, {\it {Stochastic quantization}},  {\em Phys.
  Rept.} {\bf 152} (1987) 227.

\bibitem{Parisi:1984cs}
G.~Parisi, {\it {On complex probabilities}},  {\em Phys. Lett.} {\bf B131}
  (1983) 393--395.

\bibitem{Klauder:1983sp}
J.~R. Klauder, {\it {Coherent state Langevin equations for canonical quantum
  systems with applications to the quantized Hall effect}},  {\em Phys. Rev.}
  {\bf A29} (1984) 2036--2047.

\bibitem{Aarts:2009uq}
G.~Aarts, E.~Seiler, and I.-O. Stamatescu, {\it {The complex Langevin method:
  When can it be trusted?}},  {\em Phys. Rev.} {\bf D81} (2010) 054508,
  [\href{http://arxiv.org/abs/0912.3360}{{\tt arXiv:0912.3360}}].

\bibitem{Aarts:2011ax}
G.~Aarts, F.~A. James, E.~Seiler, and I.-O. Stamatescu, {\it {Complex Langevin:
  etiology and diagnostics of its main problem}},  {\em Eur. Phys. J.} {\bf
  C71} (2011) 1756, [\href{http://arxiv.org/abs/1101.3270}{{\tt
  arXiv:1101.3270}}].

\bibitem{Seiler:2012wz}
E.~Seiler, D.~Sexty, and I.-O. Stamatescu, {\it {Gauge cooling in complex
  Langevin for QCD with heavy quarks}},  {\em Phys.Lett.} {\bf B723} (2013)
  213--216, [\href{http://arxiv.org/abs/1211.3709}{{\tt arXiv:1211.3709}}].

\bibitem{Sexty:2013ica}
D.~Sexty, {\it {Simulating full QCD at nonzero density using the complex
  Langevin equation}},  {\em Phys.Lett.} {\bf B729} (2014) 108--111,
  [\href{http://arxiv.org/abs/1307.7748}{{\tt arXiv:1307.7748}}].

\bibitem{Fodor:2015doa}
Z.~Fodor, S.~D. Katz, D.~Sexty, and C.~Torok, {\it {Complex Langevin dynamics
  for dynamical QCD at nonzero chemical potential: a comparison with
  multi-parameter reweighting}},  \href{http://arxiv.org/abs/1508.05260}{{\tt
  arXiv:1508.05260}}.

\bibitem{Nagata:2015uga}
K.~Nagata, J.~Nishimura, and S.~Shimasaki, {\it {Justification of the complex
  Langevin method with the gauge cooling procedure}},
  \href{http://arxiv.org/abs/1508.02377}{{\tt arXiv:1508.02377}},
to be published in PTEP.



\bibitem{Tsutsui:2015tua}
S.~Tsutsui and T.~M. Doi, {\it {An improvement in complex Langevin dynamics
  from a view point of Lefschetz thimbles}},
  \href{http://arxiv.org/abs/1508.04231}{{\tt arXiv:1508.04231}}.

\bibitem{Bloch:2015coa}
J.~Bloch, J.~Mahr, and S.~Schmalzbauer, {\it {Complex Langevin in
  low-dimensional QCD: the good and the not-so-good}},  in {\em {Proceedings,
  33rd International Symposium on Lattice Field Theory (Lattice 2015)}}, 2015.
\newblock \href{http://arxiv.org/abs/1508.05252}{{\tt arXiv:1508.05252}}.

\bibitem{Sinclair:2015kva}
D.~K. Sinclair and J.~B. Kogut, {\it {Exploring complex-Langevin methods for
  finite-density QCD}},  in {\em {Proceedings, 33rd International Symposium on
  Lattice Field Theory (Lattice 2015)}}, 2015.
\newblock \href{http://arxiv.org/abs/1510.06367}{{\tt arXiv:1510.06367}}.

\bibitem{Aarts:2015hnb}
G.~Aarts, F.~Attanasio, B.~J\"ager, E.~Seiler, D.~Sexty, and I.-O. Stamatescu,
  {\it {Towards the heavy dense QCD phase diagram using complex Langevin
  simulations}},  in {\em {Proceedings, 33rd International Symposium on Lattice
  Field Theory (Lattice 2015)}}, 2015.
\newblock \href{http://arxiv.org/abs/1510.09098}{{\tt arXiv:1510.09098}}.

\bibitem{Aarts:2015yuz}
G.~Aarts, F.~Attanasio, B.~J\"ager, E.~Seiler, D.~Sexty, and I.-O. Stamatescu,
  {\it {Insights into the heavy dense QCD phase diagram using Complex Langevin
  simulations}},  in {\em {Proceedings, 33rd International Symposium on Lattice
  Field Theory (Lattice 2015)}}, 2015.
\newblock \href{http://arxiv.org/abs/1510.09100}{{\tt arXiv:1510.09100}}.

\bibitem{Hayata:2015lzj}
T.~Hayata, Y.~Hidaka, and Y.~Tanizaki, {\it {Complex saddle points and the sign
  problem in complex Langevin simulation}},
  \href{http://arxiv.org/abs/1511.02437}{{\tt arXiv:1511.02437}}.

\bibitem{Mollgaard:2013qra}
A.~Mollgaard and K.~Splittorff, {\it {Complex Langevin dynamics for chiral
  Random Matrix Theory}},  {\em Phys.Rev.} {\bf D88} (2013), no.~11 116007,
  [\href{http://arxiv.org/abs/1309.4335}{{\tt arXiv:1309.4335}}].

\bibitem{Greensite:2014cxa}
J.~Greensite, {\it {Comparison of complex Langevin and mean field methods
  applied to effective Polyakov line models}},  {\em Phys. Rev.} {\bf D90}
  (2014), no.~11 114507, [\href{http://arxiv.org/abs/1406.4558}{{\tt
  arXiv:1406.4558}}].

\bibitem{Nishimura:2015pba}
J.~Nishimura and S.~Shimasaki, {\it {New insights into the problem with a
  singular drift term in the complex Langevin method}},  {\em Phys. Rev.} {\bf
  D92} (2015), no.~1 011501, [\href{http://arxiv.org/abs/1504.08359}{{\tt
  arXiv:1504.08359}}].

\bibitem{Mollgaard:2014mga}
A.~Mollgaard and K.~Splittorff, {\it {Full simulation of chiral random matrix
  theory at nonzero chemical potential by complex Langevin}},  {\em Phys.Rev.}
  {\bf D91} (2015), no.~3 036007, [\href{http://arxiv.org/abs/1412.2729}{{\tt
  arXiv:1412.2729}}].

\bibitem{Splittorff:2014zca}
  K.~Splittorff,
\emph{Dirac spectrum in complex Langevin simulations of QCD},
\emph{Phys.\ Rev.} {\bf D91} (2015), no.~3 034507,
[\href{http://arxiv.org/abs/1412.0502}{{\tt
  arXiv:1412.0502}}].

\bibitem{Nagata:2015future}
K.~Nagata, J.~Nishimura, and S.~Shimasaki, in preparation.

\end{thebibliography}


\providecommand{\href}[2]{#2}\begingroup\raggedright

\end{document}